\documentclass[twocolumn,prl,nofootinbib]{revtex4}

\usepackage{graphicx}
\usepackage{amssymb}
\usepackage{feynmf}	
\usepackage{xcolor}	

\def\be{\begin{equation}}
\def\ee{\end{equation}}
\def\ba{\begin{eqnarray}}
\def\ea{\end{eqnarray}}
\def\beq{\begin{eqnarray}}
\def\eeq{\end{eqnarray}}

\def\d{\mathrm{d}}
\def\p{{\cal P}}

\def\L*{{\cal L}_*}
\def\L{\mathcal{L}}
\def\U{\mathcal{U}}
\def\({\left(}
\def\){\right)}
\def\ie{{\it i.e. }}

\def\nn{\nonumber}
\def\p{\partial}
\def\mn{_{\mu \nu}}

\def\p{\partial}

\def\<{\langle}
\def\>{\rangle}

\newcommand{\eqref}[1]{(\ref{#1})}

\def\cs2{c_{s}^{2}}

 \def\ep{\varepsilon}

 \def\La{\Lambda}

 \def\p{\partial}

 \def\be   {\begin{equation}}   \def\ee   {\end{equation}}
 \def\bea  {\begin{eqnarray}}   \def\eea  {\end{eqnarray}}
 \def\bean {\begin{eqnarray*}}  \def\eean {\end{eqnarray*}}

\begin{document}

\title{Galileon Duality}
\author{Claudia de Rham}
\author{Matteo Fasiello}
\author{Andrew J. Tolley}
\affiliation{CERCA \& Department of Physics, Case Western Reserve University, 10900 Euclid Ave, Cleveland, OH 44106, USA}

\begin{abstract}
We show that every Galileon theory admits a dual formulation as a Galileon theory with new operator coefficients. In $n$ dimensions a free scalar field in Minkowski spacetime is dual to a $(n+1)$-th order Galileon theory which exhibits the Vainshtein mechanism when coupled to sources and superluminal propagation even on-shell.
This demonstrates that superluminal propagation at low energies is compatible  with an analytic S-matrix and causality. For point sources, the duality interchanges the strongly coupled Vainshtein regime with the weakly coupled asymptotic regime.
\end{abstract}
\maketitle

\section{Introduction}

Dualities, classical and quantum equivalences between naively distinct theories, are powerful tools in field theory as they can be useful probes of the non-perturbative structure of a theory. The simplest duality is the electric-magnetic duality of Heaviside. This was generalized in the pioneering work of Montonen and Olive \cite{Montonen:1977sn} to conjectured dualities for non-Abelian gauge theories which relate weak and strong coupling regimes, and by extension particles to solitons. Dualities also form the basis of the modern understanding of string theory.

In this \textit{letter} we demonstrate that the recently proposed Galileon models \cite{Nicolis:2008in} which arise naturally in the context of massive gravity theories \cite{deRham:2010ik,deRham:2010kj} (see also \cite{deRham:2014zqa} for a recent review) exhibit a non-trivial duality mapping Galileon theories into themselves. In particular, a ghost-free Galileon theory with a Vainshtein mechanism \cite{Vain} in place will map in general to a dual ghost-free Galileon theory which exhibits the dual analogue of the Vainshtein mechanism. As an extreme example, we show that in any dimension a free massless scalar field is equivalent to a Galileon theory with nonzero coefficients.

A Galileon is a scalar field $\pi(x)$ whose action is invariant under the global nonlinearly realized symmetry $\pi(x) \rightarrow \pi(x) + v_{\mu} x^{\mu}$. Galileons are the simplest field theories which exhibit the Vainshtein mechanism \cite{Vain}. The duality map can be defined as follows: given a Galileon $\pi(x)$, we can define the coordinate transformation
\be
\label{def1}
\tilde x^{\mu} = \Phi^{\mu}(x)=x^{\mu} + {\partial^{\mu}\pi(x)} \, .
\ee
As shown in \cite{BiHiguchi}, this transformation has an inverse $Z^a(\Phi(x))=x^a$,
\be\label{def2}
x^{\mu} =Z^{\mu}(\tilde x)= \tilde x^{\mu} + {\tilde \partial^{\mu}\rho(\tilde  x)} \,
\ee
which defines the dual Galileon $\rho(\tilde x)$. These relations are equivalent to a Legendre transformation \cite{Curtright:2012gx} and can be written as the {\it invertible} maps
\ba
&& \pi(x) = - \rho(\tilde x) - ( \tilde \partial_{\mu} \rho(\tilde x))^2 \, \\
&& \rho(\tilde x) = - \pi(x) - (  \partial_{\mu} \pi(x))^2 \, .
\ea
The map can also be thought of as a (field-dependent) diffeomorphism, which would then suggest that it is valid at the quantum level. See Ref.~\cite{deRham:2014lqa} for more details.

To cubic order, the map  is given in Ref.~\cite{BiHiguchi}. In configurations which only depend on $r^2=\sum_{\mu=0}^{d-1} s_\mu \eta_{\mu\nu} x^\mu x^\nu$, with $s_\mu=0$ or $1$, (such configurations include the static spherically symmetric one), then
\ba
\rho(r) = -\pi(r) + \sum_{n=2}^\infty\frac{(-1)^n}{n!}\p^{n-2}_r [\pi'(r)^n]\,.
\ea
The map is always invertible provided the eigenvalues of $\partial^{\mu} \p_{\nu} \pi$ and $\p^{\mu} \p_{\nu} \rho$ are greater than $-1$ which we impose as a constraint classically and quantum mechanically. This condition is analogous to imposing the requirement that the signature of the metric is $-+++$ in the Lorentzian path integral for quantum gravity  or that the determinant of the metric in GR never vanishes so that the metric is invertible. Furthermore it is natural in the context of bigravity models where a singularity develops if the metric product $f^{-1}g$ is not invertible, which amounts in the decoupling limit to requiring that the eigenvalues of $\partial^{\mu} \p_{\nu} \pi$  are greater than $-1$. Restricting ourselves to a theory where the map is invertible is therefore not a modification of the original theory but rather a natural choice for any gravitational theory and is very similar to what is performed in GR.
In what follows, we show that the dual Galileon theory is identical to a Galileon theory with  generically
 different operator coefficients.

\section{Duality Relations}

The Jacobian
of the transformation from $x$ to $\tilde x$ is
\ba
\left|\frac{\delta \tilde x^a}{\delta x^b}\right|=|\det\(\eta\mn+\Pi\mn(x)\)| \equiv |\eta+\Pi(x)|\,.
\ea

We work in a $d$-dimensional flat spacetime. For a tensor $X\mn$, we define the Lagrangian density,
\ba
\mathcal{U}_n[X]=\Lambda^{2\sigma}\ep^{\mu_1 \cdots \mu_d}\ep^{\nu_1 \cdots \nu_d}
\prod_{j=1}^n X_{\mu_j\nu_j}\prod_{k=n+1}^d \eta_{\mu_k\nu_k}
\,,
\ea
where $\ep$ is the Levi-Civita symbol. $\Lambda$ represents the strong coupling scale, or the scale at which the Galileon interactions become important, and the power $\sigma$ depends on the number of dimensions, $\sigma=1+d/2$, in particular in  $d=4$ spacetime dimensions, $\sigma=3$.

The different Galileon interactions \cite{Nicolis:2008in} for the respective fields $\rho$ and $\pi$ are expressed as $\L_n[\pi]=\pi \, \U_{n-1}[\Pi] $ and $\L_n[\rho]=\rho \, \U_{n-1}[\Sigma]$,
with the second derivative matrices $\Pi\mn\equiv \p_\mu\p_\nu \pi$ and $\Sigma\mn\equiv \p_\mu\p_\nu \rho$. The canonically normalized fields are given in terms of $\rho$ and $\pi$ as $\rho=\hat \rho/\Lambda^\sigma$ and $\pi=\hat \pi/\Lambda^\sigma$.

We now collect a number of useful mathematical identities from \cite{BiHiguchi} and use the notation introduced therein. Out of the spacetime coordinate $x^a$, let us define the variable $\tilde x^a$ as in Eq.~(\ref{def1}). Differentiating Eq.~(\ref{def2}), $Z^a(\tilde x) = x^a$, with respect to $x^b$ gives $[\partial^c Z^a](\tilde x ) \left[\eta_{bc}+  \partial_{b}\partial_c \pi(x) \right] = \delta^a_{b} $.
Inverting gives, in matrix notation
\be
\label{inverting}
[\partial Z](\tilde x ) = [\eta + \Sigma(\tilde x)]= [\eta+ \Pi(x)]^{-1} \, .
\ee
Similarly starting with $Z^a(\tilde x)=x^a$, we have on varying
\be
[\partial^a \delta \rho](\tilde x) + [\partial_b Z^a] \partial^b \delta \pi(x) =0 \,,
\ee
which can be rewritten as $[\eta+\Pi(x)] ^a_b[\partial^b \delta \rho](\tilde x) + \partial^a \delta \pi(x) =0 $.
Using Eq.~\eqref{inverting}, this is equivalent to
\be
\partial^a \left( \delta \rho(\tilde x) + \delta \pi(x) \right) =0 \, ,
\ee
which integrates to give $\delta \rho(\tilde x) \equiv - \delta \pi(x)$. This relation will allow us to relate the equations of motion and the variation of the action in the dual theories.

\section{Dual Galileons}
Starting with a Galileon theory in terms of the $\pi$ variables, we have \cite{Nicolis:2008in}
\ba
\label{action1}
S=\int \d^dx\sum_{n=2}^{d+1}c_n \L_n[\pi(x)]\,.
\ea
Varying this action with respect to the field $\pi$ gives (up to boundary terms)
\ba
\delta S=\int \d^d x\(\sum_{n=1}^d(n+1)c_{n+1}\U_n[\Pi(x)]\) \delta \pi(x)\,.
\ea
Now following the same prescription as in Ref.~\cite{BiHiguchi}, using the relations $\delta \pi(x)=-\delta \rho (\tilde x)$, $(\eta+\Pi(x))=(\eta+\Sigma(\tilde x))^{-1}$ and
$\d^d x= |\eta+\Pi(x)|^{-1} \d^d \tilde x= |\eta+\Sigma(\tilde x)| \d^d \tilde x$, the variation can be expressed in terms of $\rho$ as
\ba
\delta S=-\int \d^d \tilde x \, |\eta+\Sigma(\tilde x)| \sum_{n=2}^{d+1}nc_{n}\U_{n-1}\left[\frac{-\Sigma(\tilde x)}{\eta+\Sigma(\tilde x)}\right] \delta \rho(\tilde x)\,, \nn
\ea
which follows from varying the following action expressed entirely in terms of $\rho$,
\ba
\label{action2}
\hspace{-10pt} S_{\rm dual}\hspace{-3pt}=\hspace{-3pt}\int \d^d \tilde x \sum_{n=2}^{d+1}p_n \L_n[\rho(\tilde x)]
\equiv\hspace{-3pt}\int \d^d x \sum_{n=2}^{d+1}p_n \L_n[\rho(x)]\,,
\ea
where the Galileon coefficients $p_n$ in the dual theory are expressed in terms of the ones in the original theory as follows
\ba
\label{coefficients}
p_n=\frac{1}{n}\sum_{k=2}^{d+1}(-1)^{k} c_{k}\frac{k(d-k+1)!}{(n-k)!(d-n+1)!}\,.
\ea
Here $n! = \Gamma(n+1)$ which defines this expression for negative integers, in particular $1/(-|n|)!=0$ for $n\in \mathbb{N}_\star$.
We have thus proven the equivalence at the level of the action between the two Galileon theories defined by \eqref{action1} and \eqref{action2}.

In particular, working in four dimensions, we have
$p_2=c_2$,
$p_3=2 c_2-c_3$,
$p_4=\frac{3}{2}c_2-\frac 32 c_3+c_4$
and $p_5=\frac15\(2 c_2-3c_3+4c_4-5c_5\)$.
%
In what follows we may set $c_2\equiv p_2\equiv -1/12$ which corresponds to the proper canonical normalization for the scalar field in both representations.

\section{Galileon Transformations}
Both actions \eqref{action1} and \eqref{action2} are invariant under a nonlinearly realized global Galilean transformation \cite{Nicolis:2008in}
\be
\pi(x) \rightarrow \pi'(x) = u_{\mu} x^{\mu} + \pi(x^b) \, .
\ee
Under this transformation $\rho(x)$ transforms as a combination of a Galileon transformation and a translation
\be
\label{rhotrans}
\rho(x) \rightarrow \rho'(x) = - u_a x^a + \rho(x^b - {u^b}) \, .
\ee
To prove this, let us consider an infinitesimal transformation $\delta \pi(x) = u_{\mu}x^{\mu}$
and consider the defining relation $Z^a(\tilde x)=Z^a(x+ \partial \pi) =x^a $.
Perturbing we have
\ba
&&\left[\delta Z^a\right] (\tilde x) + [\partial_b Z^a](\tilde x)  \,  \partial^b \delta \pi(x) \, \nn \\
&& =\left[ \delta Z^a \right](\tilde x) + [\partial_b Z^a](\tilde x)  \,  u^b =0 \, .
\ea
Remembering that $Z^a(\tilde x) = \tilde x^a + \tilde \partial^a \rho(\tilde x)$, this becomes
\ba
&&[\partial^a \delta \rho](\tilde x)+ \left[\delta^a_b + \partial^a \partial_b \rho \right](\tilde x) u^b =0 \, \\
\Rightarrow && {\partial^a \delta \rho}(x)+[ \delta^a_b + \partial^a \partial_b] \rho(x) u^b =0 \,.
\ea
Integrating gives $\delta \rho(x) = - u_a x^a -u^a \partial_a \rho(x)$, which is the infinitesimal form of Eq.~\eqref{rhotrans}.
In addition, both actions \eqref{action1} and \eqref{action2} are invariant under the independent transformation
\ba
\pi(x) &\rightarrow& \pi'(x) = v_{\mu} x^{\mu} + \pi(x^b + {v^b}) \\
\rho(x) &\rightarrow& \rho'(x) =  -v_a x^a + \rho(x^b)\, .
\ea
The fundamental reason the Galileon and translation symmetries are mixing is clear in their origin in the decoupling limit of bigravity \cite{Hassan:2011zd} derived in \cite{BiHiguchi}. A Galileon transformation of $\pi$ amounts to a translation of the $f$ metric coordinates $\tilde x$, whereas a Galilean transformation of $\rho$ corresponds to a translation of the $g$ metric coordinates $x$. The duality between the dual Galileon variables $\pi(x)$ and $\rho(x)$ is related to the duality of bigravity theories under interchange of the two metrics $g$ and $f$ \cite{Hassan:2011zd,BiHiguchi}.

\section{Dual Coupling to Matter}

\subsection{Point-like source}

Although the duality may appear not to be local, we will see in what follows that not only  is it invertible and maps a local Galileon to a local Galileon but  the coupling to matter also maps in a local way.
To see this, consider the minimal coupling of $\pi(x)$ with an external field $J_{\pi}(x)$ through an interaction $S_{\rm int} = \int \d^d x J_{\pi}(x) \pi(x)$.
Following the duality map this is equivalent to
\ba
\nn
S_{\rm int} =- \int \d^d \tilde x \Bigg[|\eta+\Sigma(\tilde x)| \left( \rho(\tilde x) + ( \tilde \partial_{\mu} \rho(\tilde x))^2 \right) \\
\times\ J_{\pi}(\tilde x+ \tilde \partial \rho(\tilde x))\Bigg]\,.
\ea
Thus if the source for $\pi(x)$ is a delta function $J_{\pi}(x) = J_0 \delta^{d}(x-\bar x)$, then the same is true for the source for $\rho(x)$ with the only difference being that the source is localized at $\tilde x+ \tilde \partial \rho(\tilde x)= \bar x$.

\subsection{New Vainshtein realization}

The duality acts on the coupling to matter by making the source dependent on the Galileon. For instance, beginning with a minimal coupling of $\pi(x)$ to an external source $J_{\pi}(x)$ through $\int \d^4 x\, \pi(x) J_{\pi}(x)$, and following the same procedure as above, this is dual to a source in the equations of motion for $\rho$ which takes the form $J_{\rho} =|\eta + \Sigma| J_{\pi}(x+ \partial \rho)$. Thus the duality gives an independent means of realizing the Vainshtein mechanism through direct non-minimal coupling of the Galileon to matter, rather than through non-minimal kinetic self-interactions. Through these couplings, even a free scalar field can exhibit the Vainshtein mechanism as is implied by the duality.

\subsection{Matter fields}

In the previous example we consider an external source. From this argument one might be lead to think that the duality can map a local source located at $x$ to a non-local one. This an artifact of considering an external source with no dynamics and the same conclusion would also occur in GR after a change of coordinates for the metric had we considered an external source (see also ref.~\cite{Creminelli:2014zxa} for a recent discussion of this point).

To consider the coupling to matter more seriously, one  ought to include a coupling to dynamical degrees of freedom, for instance $S_{\rm int}=\int \d^d x \pi (x) \chi(x)$ or more generally
$S_{\rm int}=\int \d^d x \, \L(\chi(x), \p \chi(x), \pi(x), \p \pi(x))$ where $\chi$ is a dynamical scalar field with its own dynamics, for example $S_{\chi}=\int \d^d x (-\frac 12 (\p \chi)^2-V(\chi))$.

Under the duality map all dynamical fields transform. As explained in \cite{deRham:2014lqa}, the duality map can be seen as a (field dependent) diffeomorphism under which all matter fields transform under their appropriate representation. In the previous example, the scalar field $\chi(x)$ should transform as a scalar under the duality map, {\ie}as
\be
\tilde \chi (\tilde x) = \chi(x) \, .
\ee
So these local couplings to matter map to local ones in the dual representation
\ba
&&S_{\rm int}=\int \d^d x \L\left[\chi(x), \p \chi(x), \pi(x), \p \pi(x)\right] \\
\to && \tilde S_{\rm int }= \int \d^d \tilde  x |\eta+\Sigma(\tilde x)|
\tilde\L\,,
\nn
\ea
with
\ba
\tilde\L
=\L\left[\tilde\chi(\tilde x), [\eta+\Sigma(\tilde x)]^{-1} \tilde \p \tilde \chi(\tilde x), \rho(\tilde x) + ( \tilde \partial_{\mu} \rho(\tilde x))^2, \tilde \p \rho(\tilde x)\right]\,.\nn
\ea
Thus although the duality map appears to be non-local, it maps a local interacting theory into a local interacting theory. This is discussed in more depth in \cite{deRham:2014lqa}.

\section{Dual of a Free Theory}

Using Eq.~(\ref{coefficients}), we see that in any dimension, the dual of a free theory ($c_2 \ne 0 $, $c_n=0$ if $n>2$), is equivalent to a Galileon theory with nonzero coefficients for all the operators. In four dimensions the specific quintic Galileon theory with $p_2=-1/12$, $p_3=-1/6$, $p_4=-1/8$ and $p_5=-1/30$ (which is stable and does exhibit a Vainshtein mechanism about a static spherically symmetric source, {\it with  (classical) superluminal propagation} \cite{Nicolis:2008in}) is dual to a {\it free} theory $S=\int \d^4x \(-\frac 12 (\p \pi)^2\)$. The existence of a map between a naively superluminal theory and a luminal one was observed in an analogous context for conformal Galileons in \cite{Creminelli:2013fxa}.

The fact that a theory which admits classical superluminal propagation can be dual to a free theory should not come as a surprise. Furthermore these superluminalities do not imply a violation of relativistic causality (often called microcausality).
Speed of propagations and finite frequency phase and group velocities do not need to be invariant under different representations. However, the causal structure has to remain the same in both representations. In what follows we merely illustrate the fact that low energy (classical) superluminal propagation can exist in some field representations and do not necessarily imply acausality.  Superluminal group velocities have been observed in nature and are therefore clearly acceptable and do not violate causality. Causality is determined by the front velocity which should be the same in both representations. This distinction is discussed in more detail in \cite{deRham:2014lqa}.
We conjecture that a proper calculation of the front velocity (which has to include quantum corrections) would lead to a luminal result in both sides of the map.

\subsection{Plane Wave Solutions}

We shall demonstrate that the dual formulation of a free field theory, whose S-matrix is necessarily analytic and trivial, exhibits classical superluminal propagation even on-shell, in the absence of sources.
We focus on the four dimensional case which corresponds to a quintic Galileon. An exact solution of the unsourced equations of motion is a plane wave traveling in the $x_1$ direction $\bar \rho(x) = F(x^-)$, where we have defined the light-cone coordinates $x^{\pm} = (x_1 \pm t)/\sqrt{2}$ so that the Minkowski metric is $\d s^2 = 2 \d x^+ \d x^- + \d  x_i^2$. This follows since the equation of motion is
\be
\U_1 [\Sigma] + 3 \U_2 [\Sigma]+3 \U_3[\Sigma]+ \U_4 [\Sigma]=0 \, ,
\ee
and $\U_n [\bar \Sigma]=0$ for $n \ge 1$ which is manifest in light-cone coordinates since the only non-zero component of $\bar \Sigma$ is $\bar \Sigma^{++}=\bar \Sigma_{--}= F''(x^-)$ and the Levi-Civita antisymmetric structure ensures that $\ep \ep \bar \Sigma^n \eta^{4-n}=0$ for all $n \ge 1$.
Linear fluctuations around this background solution $\rho(x)= \bar \rho(x) + \delta \rho(x)$, obey the equation of motion $K^{\mu \nu}(x) \partial_{\mu} \partial_{\nu} \delta \rho(x) = 0$,
where in general
\ba
K^{\mu \nu} = \frac{6}{(d-1)!} \ep^{\mu \cdots}\ep^{\nu \cdots}\sum_{n=1}^{d}n(n+1)p_{n+1}\, \bar \Sigma^{n-1}_{..}\,  \eta^{d-n}_{..}\,.\nn
\ea
For the plane wave background in four dimensions, the only non-vanishing components are
\be
K^{++} =-2F''(x^-)  \,, \ K^{+-} =1 \, \ {\rm and}\ K^{ij} = \delta^{ij}\,, 
\ee
for $i,j=2,3$.
At high energies, for which the WKB approximation is appropriate, $\delta \rho(x) \sim A e^{i k_{\mu}x^{\mu}}$ we have $K^{\mu \nu}(x) k_{\mu}k_{\nu}=0$.
For a wave travelling in the $x_1$ direction, the two independent speeds of propagation are
\be
c_s =1 \, \quad \text{and} \quad  c_s = \frac{1-F''}{1+F''}\,,
\ee
so the speed of fluctuations along one directions is superluminal, if $F''<0$, as expressed in Ref.~\cite{Evslin:2011vh}, (see also Ref.~\cite{Burrage:2011cr}). It is important to stress that this solution is well within the regime of the effective theory, in particular $c_s>1$ can be achieved with $F''$ as small as desired.

Since we have not included coupling to matter, this theory is exactly equivalent to a free theory at the fully quantum level, and the latter is a completely causal theory, with a trivial, analytic S-matrix.

As previously explained, this simple example comes to show how superluminalities (at least classically) can be present and yet not manifest any source of acausality since the theory is free and thus clearly causal. This proves that a quantum theory can exhibit superluminality (in the group and low-frequency phase velocity), but still be causal. There is no contradiction here since we anticipate that the front velocity, whose (sub)luminality is the true measure of relativistic causality, is identical in both duality frames. Since in one frame the theory is free, there are no quantum corrections even to off-shell correlation functions, and therefore the front velocity must be luminal. In the quintic Galileon frame, the above speed of propagation is only the low energy phase and group velocity for momenta $k\ll \Lambda$. For momenta significantly larger than $\Lambda$ quantum corrections will come to dominate the calculation of the phase velocity. If, as we conjecture, the duality is consistent also at the quantum level, then the resulting front velocity will be luminal.

\subsection{Scattering Amplitudes}

In general, if we compute the S-matrix for a generic quintic Galileon theory, one would expect to recover the same non-analyticity argument as presented in \cite{Adams:2006sv}. Yet this special quintic Galileon theory is dual to a free theory. Thus it follows that not only is its S-matrix analytic, it is exactly unity. To see this, let us compute the four and five-point functions for this theory and show that they vanish. For simplicity, we work in Euclidean space.

\noindent $\bullet$ {\bf $2\to 2$ scattering}. Both the cubic and the quartic Galileon vertices play a role in the $2\to 2$ scattering as
\ba
\mathcal{M}_{\rm 4 vt}^{\rho\rho \to \rho \rho}&=&-2 \circlearrowleft p_{12}\ p_{34}^2\\
\mathcal{M}_{\rm 3 vt}^{\rho\rho \to \rho \rho}&=&2 \circlearrowleft p_{12}^3\ p_{34}\,,
\ea
where the sum $\circlearrowleft$ is performed over all the permutations of $\{p_1, \cdots, p_4\}$.
We use the shorthand notation $p_{ij}\equiv p_i \cdot p_j$, and
assume the on-shell condition, $p_i^2=0$ for all $i=1,\cdots 4$. Using the momentum conservation relation, $p_{34}=p_{12}$, we see directly that the sum from both contributions leads to a vanishing $2 \to 2$ scattering amplitude.

At the level of the $2\to 2$ scattering amplitude, there is only one crossing symmetric Lorentz invariant combination of the momenta the amplitude can depend on, and one could always have tuned the coefficient of the quartic interactions with respect to that of the cubic interactions to cancel the  $2\to 2$ scattering amplitude. At the level of the higher order scattering amplitude on the other hand there can be several allowed Lorentz invariant combinations of the momenta that are consistent with crossing symmetry. For instance for the five point function the two combinations $\circlearrowleft p_{12}^4$ and $\circlearrowleft p_{12}^2 p_{23}^2$ are independent. Since at this level there is only one coefficient of the quintic Galileon one can play with, this means that unless a `happy accident' occurs (as we shall see) we do not expect there to be sufficient parameters to set the five point function to zero.

\noindent $\bullet$ {\bf $3\to 2$ scattering}. The $2\to 2$ scattering amplitude can always be made to vanish with the appropriate parameter for the quartic Galileon. For the five-point function, there is a priori not enough freedom to ensure the vanishing of the amplitude but as we shall see in the special case of the dual to the free theory, all three of the quintic Galileon vertices combine to lead to a vanishing scattering amplitude. The different contributions are given by
\ba
&& \hspace{-10pt} \mathcal{M}_{\rm 5 vt}^{3\to 2}=-\frac 16 \circlearrowleft
p_{12}\, p_{34}\, p_{45}\, p_{35} \, ,\nn \\
&& \hspace{-10pt} \mathcal{M}_{\rm 3\&4 vt}^{3\to 2}=-\frac 14  \circlearrowleft p_{12}\Big[(p_{13}+p_{23})\, p_{45}^2 +(p_{13}+p_{23})^2\, p_{45})\Big]\nn \, , \\
&&  \hspace{-10pt}\mathcal{M}_{\rm 3 vt}^{3\to 2}=\frac14 \circlearrowleft
p_{12}\, p_{45}\,\Big[(p_{13}+p_{23} )\,p_{45}+(p_{34}+p_{35})\, p_{12}\Big]\nn\,,
\ea
where $\circlearrowleft$ indicates the sum over all the permutations of $\{p_1, \cdots, p_5\}$.
Using once again the on-shell conditions and momentum conservation, $p_5=-(p_1+p_2+p_3+p_4)$ together with $p_{34} =-(p_{12} + p_{13}+p_{14}+p_{23} + p_{24})$, we see that all three contributions are proportional to
\ba
\bar \mathcal{M}_5&=&
(p_{14} p_{23} - p_{13} p_{24})^2 +
 2 (p_{13} + p_{14} + p_{23} + p_{24})\times \nn\\
&& (p_{14} p_{23} + p_{13} p_{24}) p_{12} + \Big[p_{13}^2 +
    p_{14}^2 + (p_{23} + p_{24})^2 \nn\\
&& + 2 p_{14} (2 p_{23} + p_{24}) +
    2 p_{13} (p_{14} + p_{23} + 2 p_{24})\Big] p_{12}^2\nn\\
&& +
 2 (p_{13} + p_{14} + p_{23} + p_{24}) p_{12}^3 + p_{12}^4\,,
\ea
with
\ba
\mathcal{M}_{\rm 5 vt}^{3\to 2}=4\bar \mathcal{M}_5\,, \
\mathcal{M}_{\rm 3\&4 vt}^{3\to 2}=12\bar \mathcal{M}_5 \, \ {\rm and }\
\mathcal{M}_{\rm 3 vt}^{3\to 2}=-16\bar \mathcal{M}_5 \nn\,,
\ea
%
and the resulting $3\to 2$ scattering amplitude is thus manifestly zero!

These arguments may be generalized to any scattering amplitude as guaranteed by the duality. In fact at tree-level one may easily argue that the S-matrices are equivalent by utilizing coherent states. The tree-level S-matrix between two coherent states is given by
\be
\langle \alpha | \hat S | \beta \rangle = e^{i S(\alpha, \beta)/\hbar} \, ,
\ee
where $S(\alpha, \beta)$ is the classical action evaluated on a classical solution of the equations of motion with boundary conditions determined by the incoming and outgoing coherent states. Since the duality transformation is a classical one, although it modifies the form of the equations of motion and the solutions, it will leave invariant the action evaluated on a given classical solution and hence leave invariant the tree-level S-matrix. Then, by means of the optical theorem, we may infer that if all the tree-amplitudes vanish then so do the loops.\vspace{-10pt}

\section{Strong-Weak Coupling Duality}\vspace{-10pt}

\subsection{Duality as a Legendre Transform}

The map between $\pi$ and $\rho$ that we have been using so far was designed to preserve the vacuum $\pi=\rho=0$. However, the map is peculiar in the sense that it maps a $\mathbb{Z}_2$ symmetric theory into a non-$\mathbb{Z}_2$ symmetric one. This may be rectified by working with redefined (dual) Galileon fields $\tilde \pi = \pi + \frac{1}{2} x^2$ and $\tilde \rho = \rho + \frac{1}{2} x^2$ and the associated $\tilde \Pi = \partial \partial \tilde \pi$ and $\tilde \Sigma = \partial \partial \tilde \rho$. We then have the $\mathbb{Z}_2$ symmetry preserving maps
\ba
\Phi(x) = \partial \tilde \pi(x) \, , \, Z(x) = \partial \tilde \rho(x) \, , \, \tilde \Sigma(\Phi) = ( \tilde \Pi(x) )^{-1} \, . \nn
\ea
In this form, the duality is equivalent to the Legendre transform presented by Curtright and Fairlie in Ref.~\cite{Curtright:2012gx}.

\subsection{From strong to weak coupling}

The vacuum state with $\langle \tilde \Pi \rangle =0$ is now mapped to a state with $\langle \tilde \Sigma \rangle =\infty$. We can view this as a map between a weakly coupled vacuum state and an infinitely strongly coupled vacuum state.  In order to see the strong-weak coupling nature of the duality, consider the example of a cubic Galileon in {\it three dimensions} with Lagrangian
\be
\La^{-5}{\cal L}_{\text{cubic}} = -\frac{1}{2} (\partial \tilde \pi)^2 +  \frac{1}{4} g \Box \tilde \pi (\partial \tilde \pi)^2 \, .
\ee
The coupling constant $g$ could be absorbed into the strong coupling scale $\Lambda$, but for the purpose of the following discussion, it will be more convenient to keep them separate.
Now using $\Phi(x) = \partial \tilde \pi(x)$ and $Z(x) = \partial \tilde \rho(x)$, the cubic Lagrangian in three dimensions is dual to a cubic Lagrangian of the form
\be
\La^{-5}{\cal L}_{\text{dual cubic}} = \frac 14 \Box \tilde \rho (\partial \tilde \rho)^2 -\frac{1}{2} g (\partial \tilde \rho)^2 \, .
\ee
Since the kinetic term now contains the coupling constant $g$, it is natural to rescale as $\tilde \rho \rightarrow \breve \rho/\sqrt{g}$ to give
\be
\La^{-5}{\cal L}_{\text{dual cubic}} = -  \frac{1}{2} (\partial \breve \rho)^2 +\frac{1}{4 g^{3/2}}  \Box \breve \rho (\partial \breve \rho)^2  \, .
\ee
As we increase the coupling $g$, the original theory becomes strongly coupled, at least from the perspective of scattering $\pi$ quanta around the $\langle \pi \rangle =0 $ vacuum. However, in the dual theory, the scattering  of $\breve \rho$ quanta about the $\langle \breve \rho \rangle =0 $ vacuum is increasingly weakly coupled. \\
Phrased differently, the Vainshtein region of the $\pi$ theory occurs when the dynamics of the cubic Galileon operator dominates over the normal kinetic term. This is dual in the $\breve \rho$ theory to when the free kinetic term dominates over the interactions. Thus the naively strongly coupled Vainshtein region (from the perspective of $\pi$ scattering) is dual to a weakly coupled region. This is consistent with the familiar understanding that deep within the Vainshtein region, where scattering of quanta in vacuum are strongly coupled, the fluctuations of $\pi$ around a sourced background and nevertheless weakly coupled.
In the present case, it is the different choice of vacuum which mimics the same effect.

\subsection{Dual to Point Sources}

To illustrate the strong-weak coupling duality, consider the Euclidean version of the three dimensional theory (\ie a static Galileon in four dimensions), and a point source minimally coupled to the cubic Galileon $\pi$ with charge $q$. The spherically symmetric solution for $\tilde \pi(x)= \tilde \pi_0(r)$ satisfies $\partial_r \left( -r^2 \tilde \pi_0'(r)+ g r \tilde \pi_0'(r)^2 \right)  =0 $,
\ie $ -r^2 \pi_0'(r)+ g r \pi_0'(r)^2 = q$. The general solution which asymptotes to zero at infinity is
\be
\tilde \pi_0(r) = \int_r^{\infty} \d r  \left( \frac{-r^2 + \sqrt{4 q g r+r^4}}{2 g r} \right) \, .
\ee
The asymptotic solution $\pi_0 \sim  q/ r$ implies $\tilde x^i = \partial_i \tilde \pi=-q x^i/r^3$. This can be inverted to give $x^i =\tilde \partial^i \tilde \rho (\tilde x)=-\sqrt{q} \tilde x^i/\tilde r^{3/2}$ which implies $\tilde \rho(x) = -2 \sqrt{q} r^{1/2}$. In other words, the asymptotic part of the solution for $\tilde \pi$ maps onto the Vainshtein region for $\tilde \rho$. Similarly the Vainshtein region for $\tilde \pi$, $\tilde \pi_0(r) \sim -2\sqrt{q/g} \sqrt{r}$, maps on the asymptotic region for $\tilde \rho$, $\tilde \rho(x) \sim q/(g r)$. The complete form of the dual solution is
\be
\tilde \rho_0(r) = \int_r^{\infty} \d r  \left( \frac{-r^2 + \sqrt{4 q  r+r^4}}{2 g r} \right) \, .
\ee
The duality interchanges the `strongly coupled' Vainshtein region close to a source with the `weakly coupled' asymptotic region. The present duality is different from the proposal of \cite{Gabadadze:2012sm}. In both cases, they may shed light on the UV completion of these theories, see also \cite{Dvali:2010jz} for related attempts.\vspace{-10pt}

\section{Path Integral Measure}
So far our consideration has been entirely classical/tree-level. One may wonder whether the duality is preserved at the quantum level. Since the duality is just a field redefinition, the S-matrix which encodes on-shell physics should be invariant. The traditional arguments for the invariance of the S-matrix under field redefinitions, as utilized in the LSZ formalism, are valid for perturbative, local and invertible field redefinitions. We will show elsewhere \cite{deRham:2014lqa} that this infinitesimal duality transformation is perturbative, local and invertible and on integrating this may be utilized to argue for invariance of the entire S-matrix. These arguments are also true provided that we ignore power law divergences which are sensitive to field redefinitions (see \cite{Burgess:1992gx} for a discussion on these subtleties).

Although the S-matrix arguments are on-shell, even working off-shell, we may at least formally define a duality invariant measure for the path integral as follows:
\ba
\int \sqrt{\mathcal{D} \pi \mathcal{D} \rho} = \int \mathcal{D} \pi  \sqrt{{\rm Det}\left[\frac{\delta \rho}{\delta \pi}\right]} = \int \mathcal{D} \rho  \sqrt{{\rm Det}\left[\frac{\delta \pi}{\delta \rho}\right]} \, . \nn
\ea
Introducing a bosonic auxiliary field $b(x)$,
\be
\sqrt{{\rm Det}\left[\frac{\delta \rho}{\delta \pi}\right]} = \int \mathcal{D} b \, e^{+i \int \d^d x \int \d^d y b(x) \frac{\delta \pi(x)}{\delta \rho(y)} b(y)/2}\,.
\ee
Using $y \rightarrow \tilde y$ and $\delta \rho(\tilde x) = - \delta \pi(x)$, we have
\be
\sqrt{{\rm Det}\left[\frac{\delta \rho}{\delta \pi}\right]} =  \int \mathcal{D} b \, e^{-i \int \d^d x |\eta + \Pi(x)| \, b(x) b(\tilde x)/2} \, .
\ee
Thus the formally duality invariant form for the path integral is
\be
Z = \int \mathcal{D} \pi \int \mathcal{D} b \, e^{i  S[\pi]-i\int \d^d x |\eta + \Pi(x)| \, b(x) b(x+ \partial \pi)/2 } \, .
\ee
Finally, we need to utilize a regulator that preserves the duality. Dimensional regularization seems to be the obvious choice since the duality map works equally well in any dimension and because it is less sensitive to field redefinitions \cite{Burgess:1992gx}.

The fact that the duality could be extended to the quantum level is also independently supported by the fact that the transformation is nothing other than a (field-dependent) diffeomorphism \cite{deRham:2014lqa}. This means that the duality could only become anomalous in a theory with diffeomorphism anomalies.

 \noindent{\bf Acknowledgements:}
CdR is supported by Department of Energy grant DE-SC0009946. AJT is supported by a Department of Energy Early Career Award. We thank Kurt Hinterbichler for useful discussions.

\end{document}